\documentclass[10pt,twocolumn,letterpaper]{article}

\usepackage{cvpr}      

\usepackage{graphicx}
\usepackage{amsmath}
\usepackage{amssymb}
\usepackage{booktabs}
\usepackage{tcolorbox}
\usepackage{multirow}
\usepackage{float}
\usepackage{adjustbox}
\usepackage{bbm}

\usepackage[pagebackref,breaklinks,colorlinks]{hyperref}

\usepackage[capitalize]{cleveref}
\crefname{section}{Sec.}{Secs.}
\Crefname{section}{Section}{Sections}
\Crefname{table}{Table}{Tables}
\crefname{table}{Tab.}{Tabs.}

\def\cvprPaperID{*****} 
\def\confName{CVPR}
\def\confYear{2022}

\newcommand{\ourwork}{MaRU}

\newcommand{\assign}[1]{}

\allowdisplaybreaks

\begin{document}

\title{\ourwork{}: A Manga Retrieval and Understanding System Connecting Vision and Language}

\author{Tom Shen\thanks{Equal Contribution}\\
Stanford University\\
United States\\
{\tt\small tomshen@stanford.edu}
\and
Violet Yao$^*$\\
Stanford University\\
United States\\
{\tt\small vyao@stanford.edu}
\and
Yixin Liu$^*$\\
Stanford University\\
United States\\
{\tt\small yixinliu@stanford.edu}
}

\maketitle

\begin{abstract} 
Manga, a widely celebrated Japanese comic art form, is renowned for its diverse narratives and distinct artistic styles. However, the inherently visual and intricate structure of Manga, which comprises images housing multiple panels, poses significant challenges for content retrieval. To address this, we present \ourwork{} (Manga Retrieval and Understanding), a multi-staged system that connects vision and language to facilitate efficient search of both dialogues and scenes within Manga frames. The architecture of \ourwork{} integrates an object detection model for identifying text and frame bounding boxes, a Vision Encoder-Decoder model for text recognition, a text encoder for embedding text, and a vision-text encoder that merges textual and visual information into a unified embedding space for scene retrieval. Rigorous evaluations reveal that \ourwork{} excels in end-to-end dialogue retrieval and exhibits promising results for scene retrieval. We release our code at \url{https://anonymous.4open.science/r/manga_reader-20F1/}.
\end{abstract}

\section{Introduction\assign{Tom}} 
\label{sec:intro}


Manga, a popular form of Japanese comic art, has a significant global cultural influence, attracting a wide audience with its varied narratives and unique art styles. However, Manga presents certain challenges for content retrieval due to its predominantly visual nature. First, Manga is usually composed of images rather than text, making conventional text-based searches ineffective. Additionally, each image in Manga often contains multiple panels, each depicting different topics and scenes. This complexity adds an extra layer of difficulty for automated visual understanding. This makes it difficult for modern search systems to work with Manga.

To address these issues, we introduce \ourwork{} (Semantic Manga Retrieval), a system that employs computer vision and natural language processing techniques to efficiently locate and understand content within individual Manga frames.

\ourwork{} takes a Manga book, represented as a list of images, and a text query from the user as inputs. The query contains specific information about a dialogue or scene that the user is trying to find. One of the key requirements for \ourwork{} is to preprocess the Manga images in a way that enables efficient retrieval. The system then produces an ordered list of Manga pages that are relevant to the user's query, ranked according to their relevance scores.

The \ourwork{} system is built as a multi-stage architecture that includes: 1) An object detection model to identify text and frame bounding boxes in Manga images, 2) A Vision Encoder-Decoder model to recognize text within the identified bounding boxes, 3) A text encoder that converts the recognized text into dense embeddings, making it easier to search, and 4) A vision-text encoder that combines text and image information into a unified embedding space used for scene retrieval.

Through rigorous evaluations, \ourwork{} demonstrates outstanding performance on end-to-end dialogue retrieval tasks and shows promising results in scene retrieval tasks. Its ability to process and comprehend the complex structures within Manga makes it an invaluable tool for both automated systems and improving accessibility for Manga content.

\section{Related Work\assign{Tom+Violet}} 


To the best of our knowledge, there is currently no existing system that incorporates multilingual natural language interfaces for both dialog and scene retrieval, along with end-to-end evaluation.

In the context of retrieving Manga content using visual elements, the work by Matsui et al. \cite{mtap_matsui_2017} proposes a sketch-based interface where users draw a sketch and the system returns the most relevant frame. However, our approach differs in that it incorporates a natural language interface for both scene and dialog search, which is more efficient for describing content when a drawing canvas is not readily available. Moreover, as Manga pages typically consist of multiple frames, it is necessary to retrieve not only the entire page but also specific parts of it. In Matsui et al.'s work, candidate regions are detected using selective search, and Edge-Oriented Histogram (EOH) features are extracted to represent the page. They compute EOH features for a query sketch and retrieve the region with the best matching EOH features using nearest neighbors. In contrast, our approach, detailed in Section \ref{section:scene_retrieval}, involves fine-tuning an object detection model, DEtection TRansformer (DETR)\cite{carion2020end}, on frame labels from the Manga109 dataset \cite{mtap_matsui_2017}. Unlike traditional methods with multiple stages, DETR enables end-to-end training. We generate dense embedding representations for each frame and textual scene description queries, compare the scene description representations to all frames, and return the most similar frame.

Regarding dialog search, it is relevant to works on text detection and recognition for Manga. Detecting text in Manga presents challenges due to atypical font styles and cluttered backgrounds. While text detection is a well-studied problem, directly applying existing methods to Manga proves to be ineffective\cite{chu2018text}. Chu et al. \cite{chu2018text} developed a Manga-specific text box detection model based on modifications to the Faster RCNN \cite{ren2015faster} framework. Their approach involves detecting candidates using a region proposal network and subsequently verifying them using a CNN. In our work, we fine-tune a DETR model \cite{carion2020end} on text box labels provided by the Manga109 dataset \cite{mtap_matsui_2017}. For text recognition, we utilize the MangaOCR project \cite{manga-ocr}, which offers a pre-trained vision encoder-decoder model fine-tuned on Manga, exhibiting robust performance in extracting text from Manga text boxes.

Expanding the scope to research at the intersection of visual and textual data, or retrieval in general, numerous insightful works exist that have influenced our approach.

One relevant research stream is Image Region Captioning, which is concerned with generating textual descriptions of various image regions, enabling text search within these descriptions for Manga page retrieval. Early work in this area, such as that by Karpathy et al. \cite{karpathy2015caption}, employed a CNN+RNN architecture. More recent developments include the BLIP system \cite{li2022blip}, which utilizes a modern vision/text transformer architecture instead of the conventional CNN and RNNs. However, the image-captioning approach have their shortcomings for retrieval tasks. In particular, the image-text interaction is limited, resulting in potential information loss from the image. Furthermore, reliance on autoregressive RNNs for captioning can negatively impact performance, and these systems generally lack the ability to comprehend dialog effectively. Note that BLIP also supports direct image retrieval without resorting to image captioning, but for the purpose of this discussion, we focus on its image captioning capabilities. This system is also ineffective for retrieving Manga directly due to the multi-frame nature of Manga.

Another key area of research involves creating shared embedding spaces for images and texts. An iconic example is OpenAI's CLIP \cite{radford2021clip}, which leverages transformer models to generate unified representations of image and text data. This allows for efficient image retrieval using Nearest Neighbor algorithms. Systems like ImageBind \cite{girdhar2023imagebind} and ALIGN \cite{jia2021scaling} have further extended this paradigm to encompass multi-modal sensory data and large scale, noisy training data, respectively. These advancements have led to remarkable zero-shot performance across diverse types of images. However, their application to Manga is limited, as a single dense embedding may struggle to encapsulate the multi-frame structure of Manga pages.

A related field involves the integration of large language models for augmenting interactions, as seen in works like MiniGPT-4 \cite{zhu2023minigpt4}. These models can understand fine-grained details of images by passing all hidden states of vision transformers to the language model, facilitating retrieval tasks through prompt engineering. Recent developments, such as PandaGPT\cite{su2023pandagpt}, have extended this capacity to multi-modal understanding. However, these approaches present challenges in retrieval tasks due to the computational infeasibility of running the large language model for each image-text pair.

Lastly, Sentence Transformers \cite{reimers2019sentencebert} refine the BERT architecture to generate embeddings that capture semantic similarity between texts, making it suitable for non-exact text search. Despite its merits, it cannot be used directly for retrieving Manga - an issue that our proposed system is designed to address. By integrating sentence transformers into our dialog retrieval process, we aim to leverage its strengths while overcoming its inherent limitations.

\section{Methods} 
\subsection{Frame \& Text Box Detection }\assign{Yixin}
\label{sec:detection-model}
To extract information from Manga pages, two critical tasks are frame detection and text detection. In this project, we utilized the DETR model\footnote{https://huggingface.co/facebook/detr-resnet-50} to accurately locate the bounding boxes for frames and text boxes on the Manga pages\cite{carion2020end}. We select DETR as our detection model due to its simplicity in architecture and ease of fine-tuning. Unlike traditional detection models that involve multi-stage pipelines (e.g., region proposal network and subsequent classification in Faster RCNN \cite{ren2015faster}), the DETR model enables end-to-end training which simplifies the fine-tuning process. The DETR model is an encoder-decoder transformer with ResNet-50 \cite{he2015resnet} backbones. It consists of two prediction heads, one for predicting class labels and the other for predicting bounding boxes. DETR model is trained using a bipartite matching loss. This involves comparing the predicted classes and bounding boxes of each of the 100 object queries to the ground truth annotations, which are padded to match the length of the object queries. Then the Hungarian matching algorithm is used to find the best possible one-to-one mapping between the queries and annotations. The loss function is a linear combination of standard cross-entropy for the classes and box loss defined as:
$$\mathcal{L}_{\text{Hungarian}}(y,\hat{y}) = \sum_{i=1}^N[-\text{log}\hat{p}_{\hat{\sigma}(i)}(c_i) + \mathbbm{1}_{\{c_i\neq\emptyset\}}\mathcal{L}_{box}(b_i, \hat{b}_{\hat{\sigma}(i)})]$$

where for element $i$ of the ground truth set, $c_i$ is the target class label and $b_i$ is the ground truth box coordinates. $\hat{p}_{\hat{\sigma}(i)}(c_i)$ is the probability of class $c_i$ of prediction with index $\hat{\sigma}(i)$, with corresponding predicted box coordinates denoted as $\hat{b}_{\hat{\sigma}(i)}$. The bounding box loss $\mathcal{L}_{box}$ in the equation above is a combination of the L1 loss and generalized IoU loss\cite{Rezatofighi_2018_CVPR}, which is defined as: 
$$\mathcal{L}_{box}(b_i, \hat{b}_{\hat{\sigma}(i)}) = \lambda_{iou}\mathcal{L}_{iou}(b_i, \hat{b}_{\hat{\sigma}(i)}) + \lambda_{L_1}||b_i -\hat{b}_{\hat{\sigma}(i)}||_1$$  The model was initially pretrained on the COCO 2017 object detection dataset and has shown competitive performance, achieving an Average Precision of 42.0. This demonstrates its ability to accurately detect and classify objects in diverse contexts, which makes it a suitable choice for our frame and text detection task in Manga pages. To achieve multilabel object detection, we assign label $1$ for frame boxes and label $2$ for text boxes. We used Pytorch\cite{NEURIPS2019_9015} and Hugging Face Trainer\cite{huggingface2023} to fine-tune the pretrained model on both text and frame annotations in our Manga109 dataset, excluding the Book Dollgun \cite{dollgun} since it was used in the end-to-end evaluation in the later section.

\subsection{Transcript Extraction\assign{Tom}}
\label{sec:transcript-extraction}

Transcript extraction in our approach is achieved in two stages. Initially, a text box detection algorithm is used to identify all text bounding boxes in a given frame, and crops are extracted based on these bounding boxes. Following this, a pretrained Vision Encoder Decoder Model is applied to the crops to extract the contained text. The architecture and pretrained weights for this process are adopted directly from the MangaOCR project \cite{manga-ocr}, signifying its invaluable contribution to our work.

The vision encoder utilizes the DeiT architecture, as detailed in \cite{touvron2021deit}. The procedure begins with the rescaling of the text crop to a 224$\times$224 image. The rescaled image is then transformed into 196 patches via a CNN with a 16$\times$16 kernel size and stride 16. Each patch corresponds to a disjunct perceptive field of 16$\times$16. These patches are subsequently fed into a transformer to generate context-aware embeddings, which are used as memory for the text decoders by cross-attention. The weights used in this process, supplied by the MangaOCR project, were fine-tuned from the pretrained weights provided by \cite{touvron2021deit}.

For the text decoding, the BERT architecture \cite{devlin2019bert} is employed, a choice made by the MangaOCR project. While BERT is typically an encoder-only transformer, in this case, the decoder has been modified to allow cross-attention and decoder mode, thereby enabling it to attend to the output from the vision encoders. The weights used in this stage, provided by MangaOCR, were originally fine-tuned from BERT-Japanese \cite{bert_japanese}.

A noteworthy advantage of our transcript extraction process is the generation of transcripts in natural reading order, which enhances the readability and accessibility of the content. While the detection model we describe in section \ref{sec:detection-model} does not inherently output bounding boxes in a naturally ordered sequence, we mitigate this issue by employing a postprocessing algorithm introduced by Kovanen et al. \cite{kovanen-manga-order}. This technique orders the position information of texts and pages on three different levels, which are then hierarchically sorted to produce the final reading order. Upon obtaining the sorted bounding boxes, we can readily generate transcript texts in the corresponding order.

\subsection{Dialog Retrieval\assign{Tom}}

For the purpose of dialog retrieval, we use the SentenceBERT architecture \cite{reimers2019sentencebert} to encode each sentence in the transcripts into a dense embedding vector. SentenceBERT shares the same architecture with BERT \cite{devlin2019bert}, but has been fine-tuned to provide semantically meaningful embeddings. By independently processing each sentence with this model, the pooled embeddings for two semantically similar sentences (such as paraphrases) yield a high cosine similarity.

These dense embeddings are saved, along with their page location, as the index for retrieval. Importantly, these embeddings are preprocessed, eliminating the need for recomputation during query time. When a query is made, we compute the embedding for the user's query and find the top-$K$ embeddings in the index that have the highest cosine similarity with the query embeddings.

The entire query operation is vectorized. All the normalized index embeddings can be represented as a matrix $A$, where each row corresponds to a normalized embedding of a sentence in the Manga book. The operation $Ax$ computes all the similarities with the query embedding $x$. While the time complexity for the query operation is linear with respect to the size of the index, in practice, each book contains at most around 10,000 sentences. Given that the operation is vectorized, the query time is very short.

As the system scales up, we can employ methods such as the FAISS library \cite{johnson2019billion} for faster nearest neighbor search, ensuring efficient retrieval even with larger indices.








\subsection{Scene Retrieval\assign{Violet}}
\label{section:scene_retrieval}
In addition to dialog retrieval, which allows users to locate a page based on a specific line of the transcript, our system also provides the capability for users to find pages using scene descriptions that focus on the visual elements.

To achieve this goal, we utilize CLIP \cite{radford2021clip}, a model trained on diverse image-text pairs with impressive zero-shot image retrieval capabilities. CLIP is trained to predict the occurrence of image-text pairs within a batch of N pairs by simultaneously training an image encoder and a text encoder. The encoders learn image and text embeddings within a shared embedding space to maximize the cosine similarity between the embeddings of the N real pairs, while minimizing the similarity with incorrect pairings.

However, our findings indicate that directly applying CLIP to Manga page images does not produce satisfactory results, as discussed in more detail in Section \ref{section:scene_retrieval_experiment}. Manga pages consist of multiple frames with irregular shapes and sizes, each featuring distinct entities and interactions. Embedding the entire page results in a representation that lacks detailed information for each frame, thereby reducing the effectiveness of scene queries specific to a particular frame. Consequently, we propose an alternative approach: training an object detection model to locate frame bounding boxes, as discussed in Section \ref{sec:detection-model}, and encoding per-frame scene embeddings instead of per-page embeddings.

For image encoding, we employ a Vision Transformer (ViT) \cite{dosovitskiy2020vit}, which divides an image into fixed-size patches, linearly embeds each patch, incorporates position embeddings, and passes the resulting sequence of vectors through a standard Transformer \cite{vaswani2017transformer} encoder. Additionally, we include a [CLS] token to represent the entire image. For text encoding, we utilize a multilingual DistilBERT \cite{sanh2019distilbert} model to support both English and Japanese queries. This model is a student model trained using Multilingual Knowledge Distillation \cite{reimer2020multilingual}, based on the original Transformer model mentioned in the CLIP paper. By leveraging parallel data, the multilingual student model aligns the vector space of the teacher model across multiple languages. Weights for both the image encoder \cite{vit-clip} and text encoder \cite{multilingual-clip} are provided by the Sentence Transformers \cite{reimers2019sentencebert} library.

Similar to dialog retrieval, we index scene embeddings along with their page location for efficient retrieval. When a query is made, we compute the cosine similarity between the query embedding and all scene embeddings, enabling the identification of the top-$k$ scene embeddings with the highest cosine similarity to the query embedding.

\subsection{Baselines\assign{Violet}} 
To the best of our knowledge, there is currently no existing system that specifically focuses on the task of dialog and scene retrieval for Manga. However, there are models that have been widely used for related tasks, such as Manga OCR for Manga transcript extraction and CLIP for text-image retrieval. In this study, we establish the baseline for dialog search by directly applying Manga OCR to each page in a book. Similarly, we establish the baseline for scene search by directly applying CLIP to each page.

\subsection{Query Set\assign{Violet}} 
\label{sec:dialog-query-set}

\paragraph{Dialog Query Set}
To quantify the performance of our dialog retrieval system and compare it to the baseline, we construct a query set using ground truth transcripts. The query set is created by sampling from the transcripts and applying heuristic filters to simulate the type of queries users might ask. We utilize a multilingual Sentence Transformer model \cite{multilingual-clip} to encode all the ground truth transcripts and compute the cosine similarity between them. We retain only those pairs with the second-largest cosine similarity less than 0.5 (excluding self-similarity) and randomly select 100 queries. The rationale behind this approach is to avoid searching for lines that appear on multiple pages or have similar paraphrases, as users typically seek distinct semantic meanings. The queries are randomly sampled from the book DollGun\cite{dollgun} and can be easily expanded in the future.

Since users may not remember the exact wording of the transcripts when using our retrieval system, we also use GPT-4 \footnote{Accessed via the OpenAI API \url{https://platform.openai.com/}} to paraphrase the queries in Japanese and create a paraphrased query set, which helps simulate real use cases more effectively. We utilize the DSP framework \cite{khattab2022demonstrate} for designing the prompt. The specific prompt we use can be found in Appendix Table \ref{table:paraphrase-prompt}. Additionally, to showcase the multilingual capabilities of our system, we provide a dialog query set in English. This is achieved by translating the queries to English using GPT-4. The specific prompt can be found in Appendix Table \ref{table:japanese-to-english}. 

\paragraph{Scene Query Set}

\begin{figure}
  \centering

  \begin{subfigure}{0.23\textwidth}
    \centering
    \includegraphics[width=\linewidth]{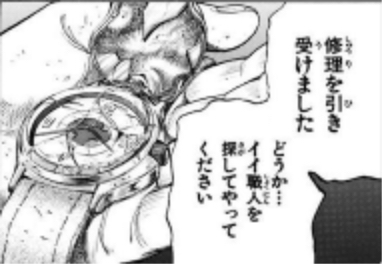}
    \caption{A hand holds a broken watch.}
    \label{fig:hand-watch}
  \end{subfigure}
  \hfill
  \begin{subfigure}{0.23\textwidth}
    \centering
    \includegraphics[width=\linewidth]{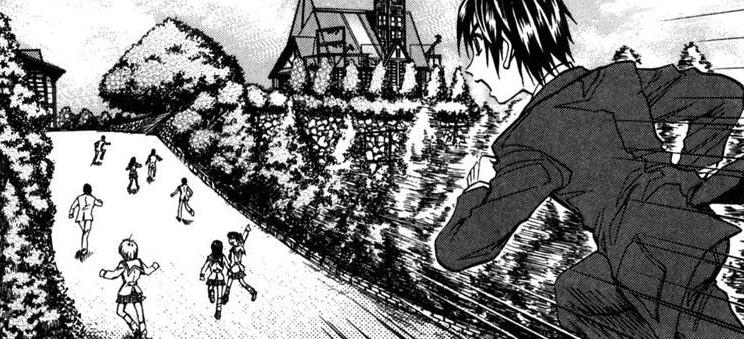}
    \caption{A group of people running on a rural path, with trees on both sides of the road.}
    \label{fig:running}
  \end{subfigure}

    \vspace{0.5cm} 

    \begin{subfigure}{0.23\textwidth}
    \centering
    \includegraphics[width=\linewidth]{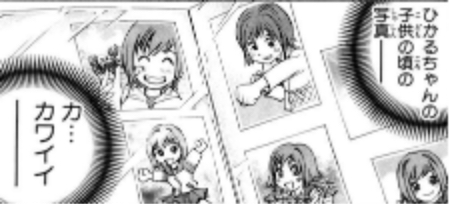}
    \caption{A photo album displays five pictures of a girl with short hair.}
    \label{fig:album}
      \end{subfigure}
  \hfill
    \begin{subfigure}{0.23\textwidth}
    \centering
    \includegraphics[width=\linewidth]{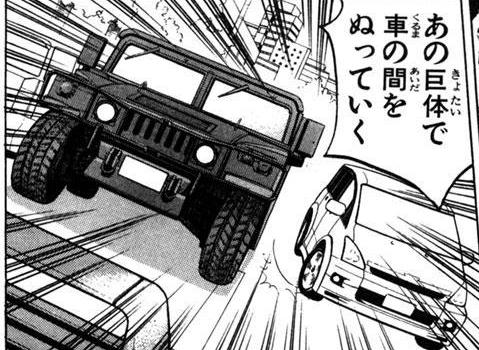}
    \caption{A jeep and a sedan are driving side by side.}
    \label{fig:jeep}
  \end{subfigure}
  \caption{Sample Scene Descriptions, \copyright Deguchi Ryusei \cite{dollgun}}
  \label{fig:scene-desc}
\end{figure}

Creating the query set for scene retrieval is more tricky. In the Manga109 dataset, there are no available ground truth scene and text description pairs that can be utilized. Additionally, directly sampling and filtering transcripts as queries may not adequately capture the essence of a scene because transcripts focus primarily on dialog and lack the illustrations and visual storytelling elements that make Manga a unique and immersive experience. Therefore, we handpick 50 frames, or scenes, predicted by the frame detection model for the book DollGun. Subsequently, we manually write 50 scene descriptions in English to simulate the types of scene descriptions users might inquire about. A few examples of these scene descriptions are shown in Figure \ref{fig:scene-desc}. To demonstrate the multilingual capabilities of our system, we include a set of scene queries in Japanese as well. The translate prompt can be found in Table \ref{table:english-to-japanese} in the appendix. To ensure accuracy and fluency, a human Japanese speaker verifies and edits these Japanese scene descriptions.



\section{Dataset} 

\begin{figure}
\centering
\includegraphics[width=0.3\textwidth]{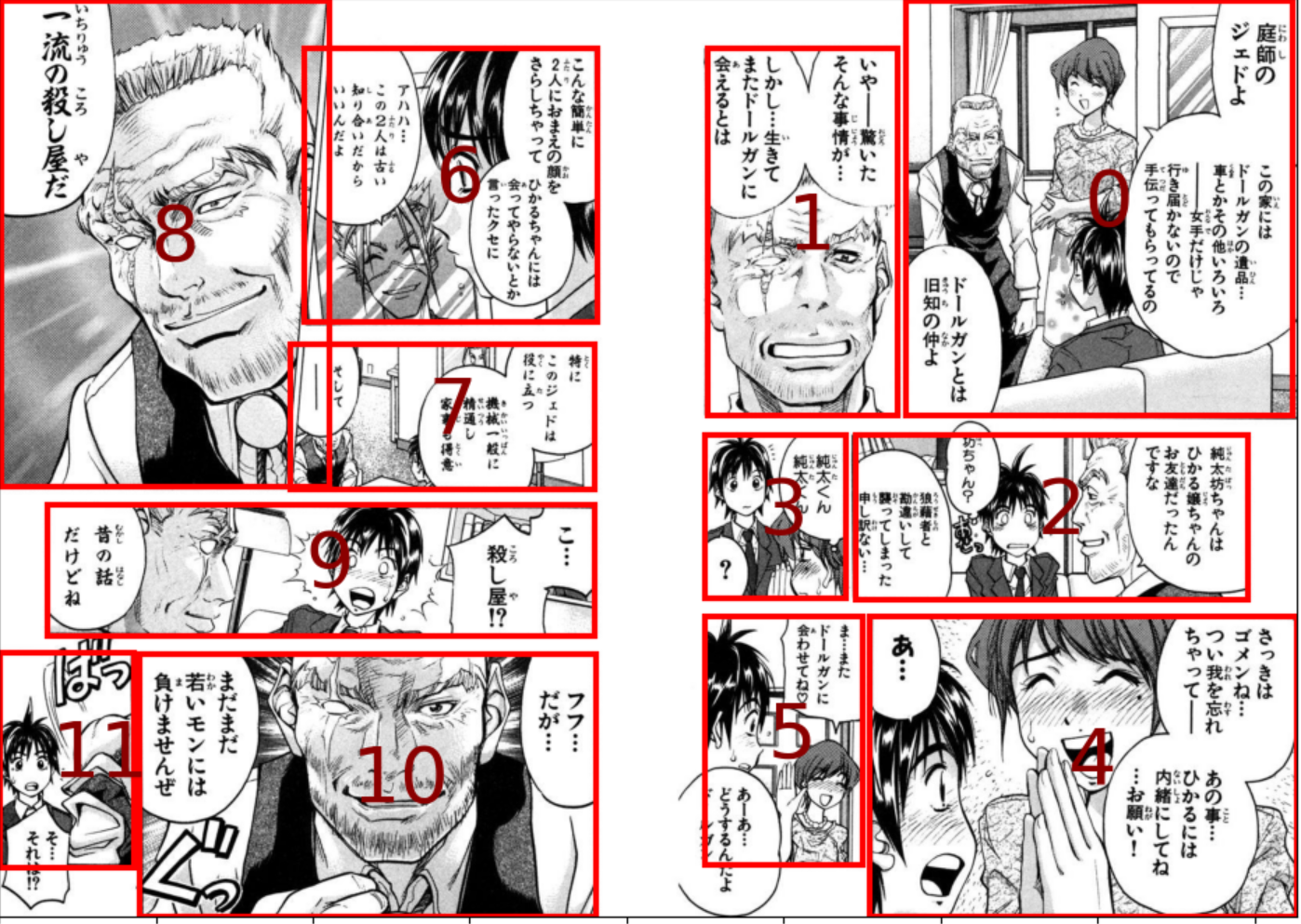}
\caption{Example Image in Manga109 Dataset with Ordered Frame Bounding Boxes, \copyright Deguchi Ryusei \cite{dollgun}}
\label{fig:dataset-example}
\end{figure}

For this project, we leverage the Manga109 dataset \cite{multimedia_aizawa_2020_manga109}, which is a compilation of 109 Manga volumes drawn by professional Manga artists in Japan, collected by the Aizawa Yamasaki Matsui Laboratory at the University of Tokyo. The dataset represents each book as a list of images, each supplemented by annotations that specify the bounding box of Manga frames (panels) and text boxes, along with corresponding text for each text box annotation. An illustration of this is provided in Figure \ref{fig:dataset-example}.

The dataset comprises a total of 10,602 pages. On average, each book encompasses around 97 pages, with a standard deviation of 18.3. We disregard book information during training and split the pages into 80\% training, 10\% validation, and 10\% testing sets, resulting in 8,482 training pages and 1,060 pages each for validation and testing. Detailed statistics are presented in Figure \ref{fig:dataset-stat} in the appendix. 

In our setup, we use images as input and employ the text and frame bounding boxes as labels for prediction. Frame detection necessitates specific data preprocessing: as the original resolution of images varies, we resize each image to either 1333$\times$800 or 800$\times$1333, depending on whether it's landscape or portrait. The images are normalized to match the mean \texttt{[0.485, 0.456, 0.406]} and standard deviation \texttt{[0.229, 0.224, 0.225]}. We also adjust bounding boxes to align with the resized images. We exclude the book DollGun\cite{dollgun} from the dataset, as it will be used for our end-to-end retrieval evaluation.

Image preprocessing is also crucial for transcript extraction and scene retrieval tasks. We resize all detected image crops to 224$\times$224. For transcript extraction, image crops are normalized to match mean 0.5 and standard deviation 0.5. In scene retrieval, we employ center cropping to maintain the scale, normalizing image crops to meet mean \texttt{[0.485, 0.456, 0.406]} and standard deviation \texttt{[0.229, 0.224, 0.225]}.


\section{Experiments} 
\subsection{Frame \& Text Box Detection\assign{Yixin}}
We fine-tune the pretrained DETR model using the AdamW optimizer with an initial learning rate of 1e-4. To facilitate the learning process, a linear learning rate scheduler was employed. We use a batch size of 8, and conduct 10 training epochs. The training and validation loss curves are shown in Figure \ref{fig:detection loss} in the appendix. We select those hyperparameters by trying different combinations, training for only a few epochs, and find the best one. Since there is almost no gap between training loss and validation loss, we do not observe any overfitting here.

 We evaluate the performance of our object detection model on the test set using two key metrics: mean average precision (mAP) and mean average recall (mAR). Average Precision (AP) is defined as area under the precision-recall curve, mAP is the average of AP across different classes. Average Recall(AR) describes the area doubled under the recall-IoU curve for IoU in [0.5, 1]. In our metrics table, the objects are categorized into three sizes: small, medium, and large, based on their area in pixels. Specifically, small objects corresponded to an area smaller than $32^2$ pixels, medium objects fell within the range of $32^2$ pixels to $96^2$ pixels, and large objects correspond to areas greater than $96^2$ pixels.
 
 We utilize multiple variations of mAP. The metrics mAP, mAP\_50, and mAP\_75 represent the mean average precision calculated under different intersection over union (IoU) thresholds. The mAP metric uses 10 IoU thresholds ranging from 0.50 to 0.95 in steps of 0.05 and returns their average, capturing precision at various thresholds. mAP\_50 uses an IoU threshold of 50\%, while mAP\_75 uses an IoU threshold of 75\%. Furthermore, we employ the metric mAR\_k, which corresponds to the mean average recall given k detections per image. This metric measures the model's ability to recall objects with varying numbers of detections per image.

\begin{table}[h]
\centering
\begin{tabular}{l|c}
Metric & Model Performance \\ \hline \hline
mAP & 0.743\\
mAP\_50 & 0.929 \\
mAP\_75 & 0.803 \\
mAP\_large & 0.799 \\
mAP\_medium & 0.277 \\
mAP\_small &0.206 \\
mAR\_1 & 0.076 \\
mAR\_10 & 0.624 \\
mAR\_100 & 0.797 \\
mAR\_large & 0.852 \\
mAR\_medium & 0.330 \\
mAR\_small & 0.350 \\
mAP (frames) & 0.878\\
mAP (text boxes) & 0.607\\
mAR\_100 (frames) & 0.912\\
mAR\_100 (text boxes) & 0.682\\

\end{tabular}
\caption{Quantitative Performance Metrics for Detection Model}
\label{tab:detection-evaluation}
\end{table}

In Table \ref{tab:detection-evaluation}, the rows representing the performance of the model are calculated for both frame detection and text detection, except for the last four rows where the metrics are evaluated per label.

The obtained results indicate a mAP of 0.743 and a mAR\_100 of 0.797, indicating that the detectors perform well in terms of precision and recall for both frame detection and text detection. It is worth noting that mAP\_large is significantly higher than mAP\_medium and mAP\_small, which is consistent with mAR\_large being much higher than mAR\_medium and mAR\_small. Since Manga pages in our dataset have a size of 1333x800 pixels, small objects account for area less than 0.1\% of the entire page, and medium objects occupy less than 0.9\% of the page. Consequently, these small frames and text boxes are less important in our retrieval task, as they are less likely to contain distinct or meaningful information that the user wants to retrieve. Regarding frames, none of the frames in our dataset falls into the small category, and the majority of them belong to the large category. Therefore, we can tolerate the detectors' relatively lower performance on small objects, given the high mAP\_large and mAR\_large values.

Additionally, the last four rows of the results indicate that frame detection outperforms text box detection in terms of mean average precision and recall. This observation can be attributed to the relatively straightforward nature of frame detection, as frames have clear boundaries and are generally larger in size compared to text boxes.

\subsection{Dialog Retrieval\assign{Tom}} 

\begin{table*}[!h]
\centering
\small
\begin{tabular}{|l|l|l|l|l|l|l|l|l|}
\hline
\multirow{2}{*}{\textbf{Method}} &
  \multicolumn{2}{c}{\textbf{k=1}} &
  \multicolumn{2}{c}{\textbf{k=5}} &
  \multicolumn{2}{c|}{\textbf{k=10}}& 
  \multicolumn{2}{c|}{\textbf{k=25}} \\
& \textbf{MRR} & \textbf{Avg Success} & \textbf{MRR} & \textbf{Avg Success} & \textbf{MRR} & \textbf{Avg Success} & \textbf{MRR} & \textbf{Avg Success} \\
\hline
\textbf{Baseline} & 0.000 & 0.000 & 0.000 & 0.000 & 0.000 & 0.000 & 0.000 & 0.000\\
\hline
\textbf{\ourwork{} End-to-End} & 0.980 & 0.980 & 0.987 & 1.000 & 0.987 & 1.000 & 0.987 & 1.000\\
\hline
\textbf{With Gold Bounding Boxes} & 0.940 & 0.940 & 0.950 & 0.960 & 0.957 & 1.000 & 0.957 & 1.000\\
\hline
\textbf{With Gold Bounding Boxes and Text} & 0.980 & 0.980 & 0.987 & 1.000 & 0.987 & 1.000 & 0.987 & 1.000\\
\hline
\end{tabular}\caption{Quantitative Performance Metrics for Text Retrieval on Paraphrased Queries}\label{table:text-metrics-paraphrased}\end{table*}

In the evaluation of our dialog retrieval system, we adopt an approach that closely mimics real-world usage scenarios. To begin with, we create a query set and select a book from the database to use as the index. The index is essentially a list of images drawn from a book, while the query set comprises a list of strings, each of which represents a query. The query set is created using the method described in \ref{sec:dialog-query-set}. 

Regarding performance metrics, we adopt two measurements appropriate for retrieval tasks, namely Mean Reciprocal Rank (MRR) at $K$ and Average Success at $K$. For MRR@$K$, we select the top $K$ relevant pages for each query using our model. The score for each query is then determined by the reciprocal of the rank of the first correct page in the retrieval result. Hence, a rank of one scores $1$, rank two scores $1/2$, rank three scores $1/3$, and so forth. Queries for which no correct page is retrieved score zero. We then compute the mean score across all queries. This metric gauges not only the success rate of each query, but also the order of the retrieval results \cite{voorhees1999proceedings}.

Average Success at $K$ operates similarly to MRR@$K$, but with a more binary scoring approach. Each query receives a score of 1 if a correct answer is within the retrieved results, and 0 otherwise. This metric is more lenient than MRR@$K$ and provides an "all or nothing" score that disregards the order of the retrieval results.

\paragraph{Quantitative Results} Our analysis focuses on evaluating the model performance using the book DollGun by Ryusho Deguchi from the Manga109 Dataset \cite{dollgun} \cite{multimedia_aizawa_2020_manga109}. This book has been excluded from training pipeline. We conducted experiments using three distinct query sets to cover a broad spectrum of practical use cases.

The ``Exact Match'' query set uses dialogue that appears verbatim in the book. The ``Paraphrased'' set takes dialogues from the book and paraphrases them utilizing OpenAI GPT-4. The paraphrased query set provides a more authentic representation of real-world scenarios, as users often recall the gist of a dialogue but not the precise wording. Finally, the ``English'' set features queries translated into English using GPT-4. This last set mirrors situations where users attempt to query a book using a language different from the original text.

Our evaluation utilizes three retriever methods. The baseline approach simply performs Optical Character Recognition (OCR) on entire pages. The ``Gold'' method, on the other hand, uses information from the dataset instead of employing an end-to-end approach. The results of these evaluation on paraphrased are displayed in Tables \ref{table:text-metrics-paraphrased}. To demonstrate the multilingual performance of our system, we also evaluate on the English query set and result is shown in Appendix Table \ref{table:text-metrics-english}. The evaluation on ``exact-match'' query is also provided in Appendix Table \ref{table:text-metrics-exact-match}.

The end-to-end model demonstrates a vastly superior performance compared to the baseline across all query sets. For instance, in the ``Exact Match" set, the baseline has a MRR and Average Success rate of 0.0 for all top-K values (1, 5, 10, 25). In contrast, the end-to-end model achieves perfect scores (1.0) for both MRR and Average Success.

Our end-to-end system's dialog retrieval performance remains impressive on the paraphrased and English query sets. This result demonstrates that the system can effectively handle queries even when the user doesn't remember the precise dialogue. For instances where the score is less than perfect, the score doesn't improve even when the ground truth bounding box and text are used for indexing. This outcome suggests that any errors might be attributed to the Sentence Transformer model we employed, rather than our text bounding box detection model.

Certainly, our evaluation does bear a certain limitation. The paraphrased or translated text used for our query sets is generated by GPT-4, which, upon inspection, retains all the details of the original text. However, in more real-world scenarios, users might forget some of the dialogue details, thereby introducing more noise into their queries compared to those generated by GPT-4. Consequently, the current performance metrics could possibly overestimate the system's efficiency in such practical situations.

\paragraph{Qualitative Analysis} The qualitative analysis conducted seeks to address the limitations previously discussed by utilizing more realistic queries. These queries may not contain the entirety of the details present in the desired dialogue and can be noisier, simulating the manner in which users might search in a real-world scenario.

\begin{figure}[!ht]
  \centering

      \begin{subfigure}{0.35\textwidth}
    \centering
    \includegraphics[width=\linewidth]{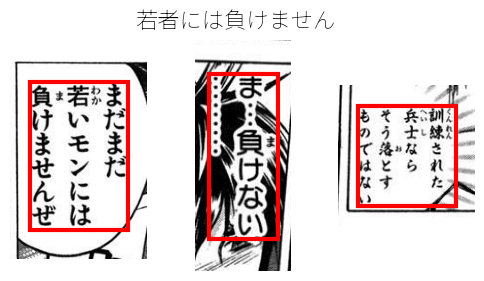}
    \caption{Query: I will not concede to the young.\\Correct (\#1) It's still too early to lose to the young guys!}
    \label{fig:text1}
  \end{subfigure}

      \begin{subfigure}{0.35\textwidth}
    \centering
      \includegraphics[width=\linewidth]{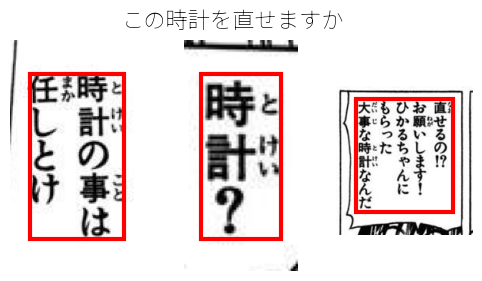}
    \caption{Query: Can you fix the clock? \\Correct (\#3) Oh you can fix it? Thanks a lot! This clock is from Hikaru and is so important to me.}
    \label{fig:text2}
      \end{subfigure}

    \begin{subfigure}{0.35\textwidth}
    \centering
      \includegraphics[width=\linewidth]{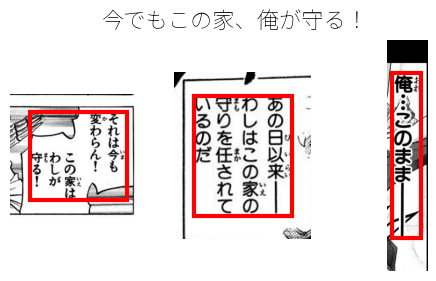}
    \caption{Query: I will protect my family at this time! \\Correct (\#1) Same as before. I will protect my family!}
    \label{fig:text3}
      \end{subfigure}

  \caption{Visualizing top 3 retrieved texts, \copyright Deguchi Ryusei \cite{dollgun}}
  \label{fig:text-desc}
\end{figure}

For the qualitative analysis, we began by indexing the book. Then, for each query, the model returned the top-3 relevant predicted bounding boxes, as shown in Figure \ref{fig:text-desc}. This approach allowed us to evaluate the model's performance across different queries and gauge its robustness.

In example \ref{fig:text1}, the query is framed in a semi-formal tone, while the correct dialogue is in a casual tone. Our model successfully retrieved the correct dialogue, demonstrating its robustness in handling queries and dialogues that differ in speaking styles and tones.

Example \ref{fig:text2} was a challenging case where the correct dialogue contained significantly more details compared to the query. Although the model did not retrieve the desired dialogue as the top-1 result, it still featured among the top-3. This indicates that the system is resilient, even when the query omits many details present in the desired dialogue.

In example \ref{fig:text3}, the desired dialogue was housed within a text bubble with an irregular shape, composed of two text boxes joined together. The successful retrieval of this dialogue showcases the capability of our text bounding box detection model to adeptly manage such cases, which contributes positively to the overall text retrieval performance.

\begin{table*}[!h]
    \centering
    \small
  \begin{tabular}{|l|l|l|l|l|l|l|l|l|}
    \hline
    \multirow{2}{*}{\textbf{Method}} &
      \multicolumn{2}{c}{\textbf{k=1}} &
      \multicolumn{2}{c}{\textbf{k=5}} &
      \multicolumn{2}{c|}{\textbf{k=10}}& 
      \multicolumn{2}{c|}{\textbf{k=25}} \\
    & \textbf{MRR} & \textbf{Avg Success} & \textbf{MRR} & \textbf{Avg Success} & \textbf{MRR} & \textbf{Avg Success} & \textbf{MRR} & \textbf{Avg Success} \\
    \hline
    \textbf{Baseline} &  0.060 & 0.060 & 0.125 & 0.240 & 0.145 & 0.380 & 0.154 & 0.540\\
    \hline
    \textbf{\ourwork{} End-to-End} & 0.320 & 0.320 & 0.413 & 0.600 & 0.429 & 0.720 & 0.435 & 0.820\\
    \hline
    \textbf{With Gold Bounding Boxes} & 0.340 & 0.340 & 0.427 & 0.580 & 0.442 & 0.700 & 0.454 & 0.860\\
    \hline
  \end{tabular}
  \caption{Quantitative Performance Metrics for Scene Retrieval on English Queries}
  \label{table:scene-metrics}
\end{table*}

\subsection{Scene Retrieval\assign{violet}}
\label{section:scene_retrieval_experiment}
\paragraph{Quantitative Results} 

In this section, we provide a quantitative analysis of our scene retrieval system, comparing its performance with the baseline and a system using a gold bounding box. Table \ref{table:scene-metrics} presents the MRR and Average Success rate for each method at different values of k.

The baseline method utilizes CLIP to directly embed each Manga page. This method yields low MRR and Average Success rates across all values of k. At k=1, the baseline method achieves an Average Success rate of 0.06. As k increases, the performance improves slightly but remains relatively low, with an MRR of 0.145 and an Average Success rate of 0.38 at k=10. Our end-to-end method, which utilizes predicted bounding boxes, significantly improves upon the baseline  across all values of k. We employ our frame detection model to predict the bounding boxes for scenes within a Manga page. At k=1, the predicted bounding box method achieves an Average Success rate of 0.32. As k increases, the performance continues to improve, reaching an MRR of 0.435 and an Average Success rate of 0.82 at k=25. The gold bounding box method utilizes the ground truth frame bounding boxes provided in the Manga109 dataset. Notably, this method achieves only slightly higher results, with an Average Success rate of 0.34 at k=1. This highlights the effectiveness of our predicted bounding box approach in capturing the frames within the Manga pages. Furthermore, to demonstrate our system's multilingual capability, we report the MRR and Average Success rate on the translated Japanese scene query set in Appendix Table \ref{table:scene-metrics-japanese}.

The bottleneck in the system lies with the CLIP model, likely due to its limited exposure to Manga content during training. Future work can involve fine-tuning CLIP with a Manga training corpus to better understand and retrieve Manga-specific scenes.

\paragraph{Qualitative Analysis}
Queries that describe specific actions or interactions between characters, such as firing a missile (see Figure \ref{fig:missile} in the appendix), holding a steering wheel, or kissing, tend to yield accurate results. Additionally, queries that mention specific visual elements or settings, such as a photo album displaying pictures of a girl (see Figure \ref{fig:album} in the appendix), a bird's eye view of a town, or hand grenades, are more likely to retrieve the desired scene. Furthermore, queries that include descriptions of characters' appearances, like a beautiful girl with wings, a boy in a suit and tie, or a man wearing sunglasses, often lead to successful scene retrieval.

However, there are also categories of queries where the system struggles. Queries involving complex interactions or relationships between multiple characters, such as a man holding scissors with others looking at him (see Figure \ref{fig:scissor} in the appendix), or broken paintings on the ground while the boy anxiously holds an injured older man, present challenges for accurate retrieval. Queries describing emotional states or expressions, like a girl anxiously shouting something, a blushing girl, or a scary man extending his hand, also prove to be more difficult for the system. Additionally, queries with ambiguous or less distinctive descriptions, such as two girls sitting on the bed and two girls standing, or a hand reaching for a cup of coffee, may result in less accurate retrieval.


Furthermore, it is worth noting that certain ``incorrect" pages can still be considered valid results as they precisely match the information described in the query. This occurs with queries that describe commonly depicted scenes, as shown in the example query ``A girl hugs a boy" (see Figure \ref{fig:hug} in the appendix). In some cases, incorrect retrieval results do include the objects mentioned in the query, but they may be missing other crucial elements. For instance, an incorrect retrieval of the query ``A man points a gun at his own head" (see Figure \ref{fig:point-head} in the appendix) includes a scene with a gun present, but the gun is not directly aimed at the character's head.

\section{Conclusion}


In conclusion, we propose \ourwork{}, a system that efficiently retrieves dialogues and scenes within Manga frames. \ourwork{} utilizes a multi-stage architecture. It employs object detection to identify text and frame bounding boxes, followed by a vision encoder-decoder model for text recognition. The recognized text is then processed using a text encoder to convert it into dense embeddings, which are indexed for dialog retrieval. Furthermore, \ourwork{} incorporates a vision-text encoder that combines textual and visual information into a unified embedding space, enabling the retrieval of relevant scenes based on user scene description queries. By introducing this work, we enhance the understanding of and improve accessibility to Manga content by comprehending its complex structures.


Future extensions to the system involve implementing character recognition algorithms for enhanced attribution and understanding of visual content. Efficient nearest neighbor search for dialog and scene retrieval can be achieved by utilizing the FAISS library \cite{johnson2019billion}. To address the bottleneck in scene retrieval, future work can fine-tune the CLIP model with a Manga training corpus for improved comprehension and retrieval of Manga-specific scenes.

\clearpage
\section{Appendices} 



\begin{table}[!b]
\begin{tcolorbox}[colback=green!5!white,colframe=green!75!black]
Please paraphrase the sentence in Japanese without using the original language.

---

Follow the following format.

Original Sentence:: \$\{the sentence to be paraphrased\} \\
Paraphrase:: \$\{the paraphrased version of the sentence\}

---

Question: [query] \\
Answer:
\end{tcolorbox}
\caption{Paraphrase Prompt}
\label{table:paraphrase-prompt}
\end{table}

\begin{table}
\begin{tcolorbox}[colback=green!5!white,colframe=green!75!black]
Please translate the sentence from English to Japanese.

---

Follow the following format.

Original Sentence in English: \$\{the sentence to be translated\} \\
Translated Sentence in Japanese: \$\{the translated version of the sentence.\}

---

Original Sentence in English: [query] \\
Translated Sentence in Japanese:
\end{tcolorbox}
\caption{Translate Prompt: English to Japanese}
\label{table:english-to-japanese}
\end{table}

\begin{table}
\begin{tcolorbox}[colback=green!5!white,colframe=green!75!black]
Please translate the sentence from Japanese to English.

---

Follow the following format.

Original Sentence in Japanese: \$\{the sentence to be translated\} \\
Translated Sentence in English: \$\{the translated version of the sentence.\}

---

Original Sentence in Japanese: [query] \\
Translated Sentence in English:
\end{tcolorbox}
\caption{Translate Prompt: Japanese to English}
\label{table:japanese-to-english}
\end{table}

\begin{table*}[!h]
\centering
\small
\begin{tabular}{|l|l|l|l|l|l|l|l|l|}
\hline
\multirow{2}{*}{\textbf{Method}} &
  \multicolumn{2}{c}{\textbf{k=1}} &
  \multicolumn{2}{c}{\textbf{k=5}} &
  \multicolumn{2}{c|}{\textbf{k=10}}& 
  \multicolumn{2}{c|}{\textbf{k=25}} \\
& \textbf{MRR} & \textbf{Avg Success} & \textbf{MRR} & \textbf{Avg Success} & \textbf{MRR} & \textbf{Avg Success} & \textbf{MRR} & \textbf{Avg Success} \\
\hline
\textbf{Baseline} & 0.000 & 0.000 & 0.000 & 0.000 & 0.000 & 0.000 & 0.000 & 0.000\\
\hline
\textbf{\ourwork{} End-to-End} & 1.000 & 1.000 & 1.000 & 1.000 & 1.000 & 1.000 & 1.000 & 1.000\\
\hline
\textbf{With Gold Bounding Boxes} & 1.000 & 1.000 & 1.000 & 1.000 & 1.000 & 1.000 & 1.000 & 1.000\\
\hline
\textbf{With Gold Bounding Boxes and Text} & 1.000 & 1.000 & 1.000 & 1.000 & 1.000 & 1.000 & 1.000 & 1.000\\
\hline
\end{tabular}\caption{Quantitative Performance Metrics for Text Retrieval on Exact Match Queries}\label{table:text-metrics-exact-match}\end{table*}
\begin{table*}[!ht]
\centering
\small
\begin{tabular}{|l|l|l|l|l|l|l|l|l|}
\hline
\multirow{2}{*}{\textbf{Method}} &
  \multicolumn{2}{c}{\textbf{k=1}} &
  \multicolumn{2}{c}{\textbf{k=5}} &
  \multicolumn{2}{c|}{\textbf{k=10}}& 
  \multicolumn{2}{c|}{\textbf{k=25}} \\
& \textbf{MRR} & \textbf{Avg Success} & \textbf{MRR} & \textbf{Avg Success} & \textbf{MRR} & \textbf{Avg Success} & \textbf{MRR} & \textbf{Avg Success} \\
\hline
\textbf{Baseline} & 0.000 & 0.000 & 0.000 & 0.000 & 0.000 & 0.000 & 0.000 & 0.000\\
\hline
\textbf{\ourwork{} End-to-End} & 1.000 & 1.000 & 1.000 & 1.000 & 1.000 & 1.000 & 1.000 & 1.000\\
\hline
\textbf{With Gold Bounding Boxes} & 0.990 & 0.990 & 0.995 & 1.000 & 0.995 & 1.000 & 0.995 & 1.000\\
\hline
\textbf{With Gold Bounding Boxes and Text} & 1.000 & 1.000 & 1.000 & 1.000 & 1.000 & 1.000 & 1.000 & 1.000\\
\hline
\end{tabular}\caption{Quantitative Performance Metrics for Text Retrieval on English Queries}\label{table:text-metrics-english}\end{table*}
\begin{table*}[!ht]
    \centering
    \small
  \begin{tabular}{|l|l|l|l|l|l|l|l|l|}
    \hline
    \multirow{2}{*}{\textbf{Method}} &
      \multicolumn{2}{c}{\textbf{k=1}} &
      \multicolumn{2}{c}{\textbf{k=5}} &
      \multicolumn{2}{c|}{\textbf{k=10}}& 
      \multicolumn{2}{c|}{\textbf{k=25}} \\
    & \textbf{MRR} & \textbf{Avg Success} & \textbf{MRR} & \textbf{Avg Success} & \textbf{MRR} & \textbf{Avg Success} & \textbf{MRR} & \textbf{Avg Success} \\
    \hline
    \textbf{Baseline} &  0.04 & 0.04 & 0.112 & 0.24 & 0.122 & 0.32 & 0.136 & 0.54\\
    \hline
    \textbf{End-to-End} & 0.22 & 0.22 & 0.314 & 0.46 & 0.329 & 0.58 & 0.339 & 0.72\\
    \hline
    \textbf{With Gold Bounding Boxes} & 0.26 & 0.26 & 0.339 & 0.46 & 0.358 & 0.6 & 0.374 & 0.84\\
    \hline
  \end{tabular}
  \caption{Quantitative Performance Metrics for Scene Retrieval on Japanese Queries}
  \label{table:scene-metrics-japanese}
\end{table*}
\begin{figure}[!h]
\centering
\includegraphics{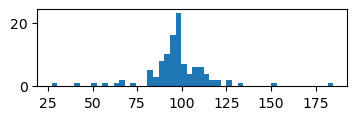}
\caption{Number of pages in books}
\label{fig:dataset-stat}
\end{figure}

\begin{figure}[h]
     \centering
     \includegraphics[width=0.4\textwidth]{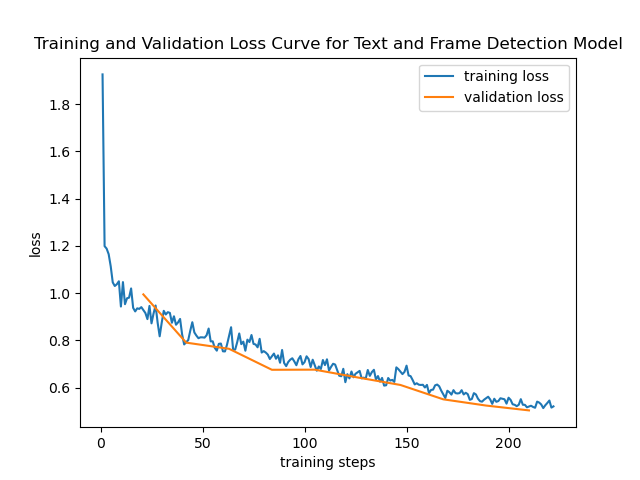}
     \caption{Training and Validation Loss for Detection Model}
     \label{fig:detection loss}
 \end{figure}

 \begin{figure*}[h]
  \centering

      \begin{subfigure}{0.85\textwidth}
    \centering
    \includegraphics[width=\linewidth]{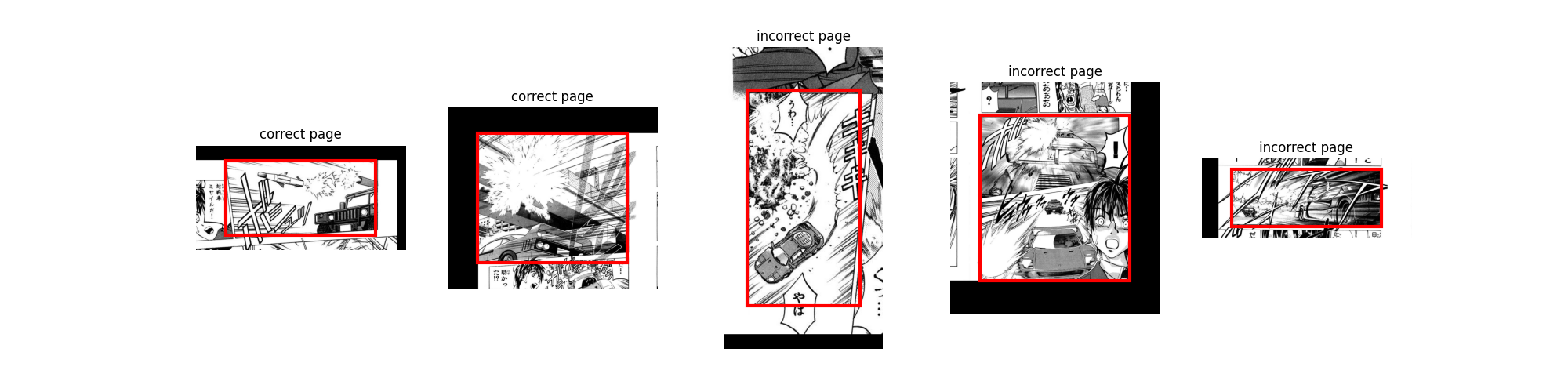}
    \caption{A car firing a missile.}
    \label{fig:missile}
  \end{subfigure}

    \vspace{0.5cm} 

    \begin{subfigure}{0.85\textwidth}
    \centering
    \includegraphics[width=\linewidth]{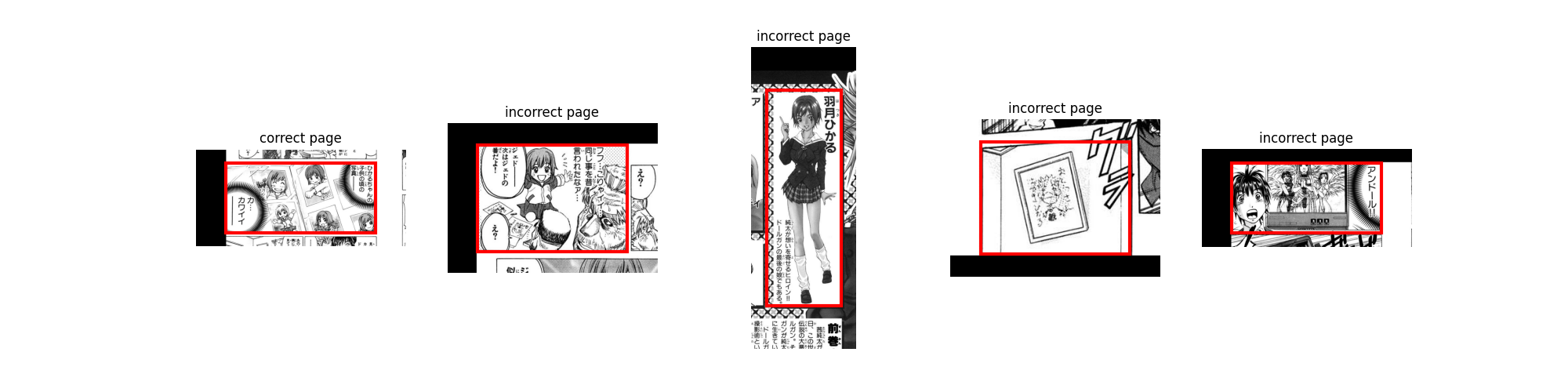}
    \caption{A photo album displays five pictures of a young girl with short hair.}
    \label{fig:album}
  \end{subfigure}

    \vspace{0.5cm} 

    \begin{subfigure}{0.85\textwidth}
    \centering
    \includegraphics[width=\linewidth]{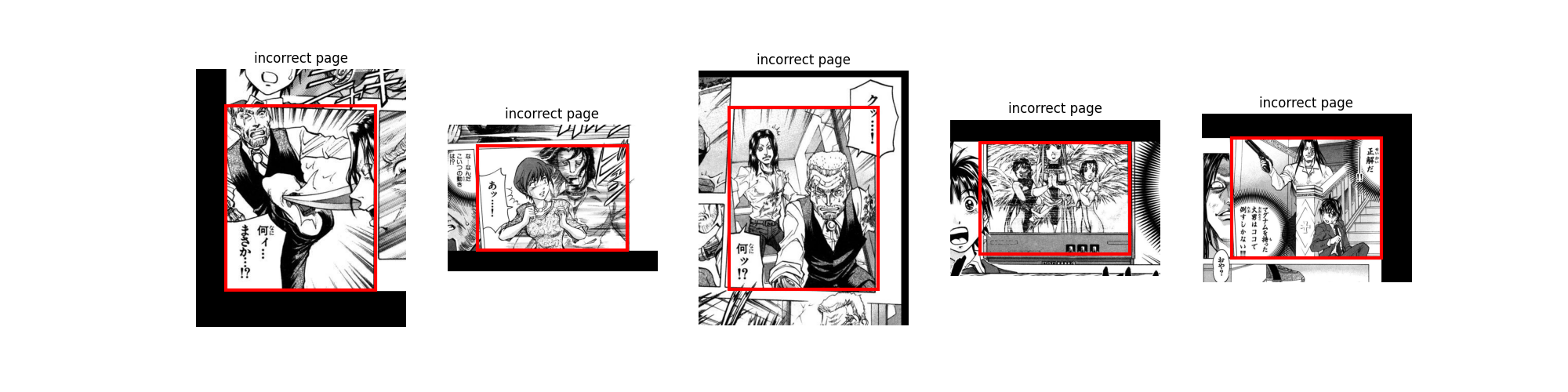}
    \caption{A man holds a pair of scissors with a man and a woman looking at him.}
    \label{fig:scissor}
  \end{subfigure}

    \vspace{0.5cm} 

  \begin{subfigure}{0.85\textwidth}
    \centering
    \includegraphics[width=\linewidth]{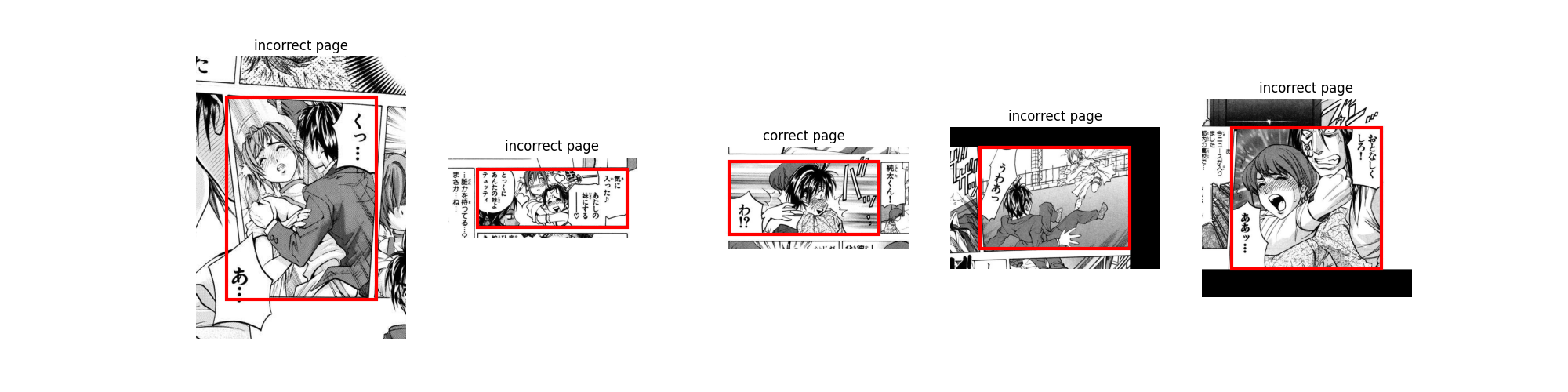}
    \caption{A girl hugs a boy.}
    \label{fig:hug}
  \end{subfigure}

    \vspace{0.5cm} 

    \begin{subfigure}{0.85\textwidth}
    \centering
    \includegraphics[width=\linewidth]{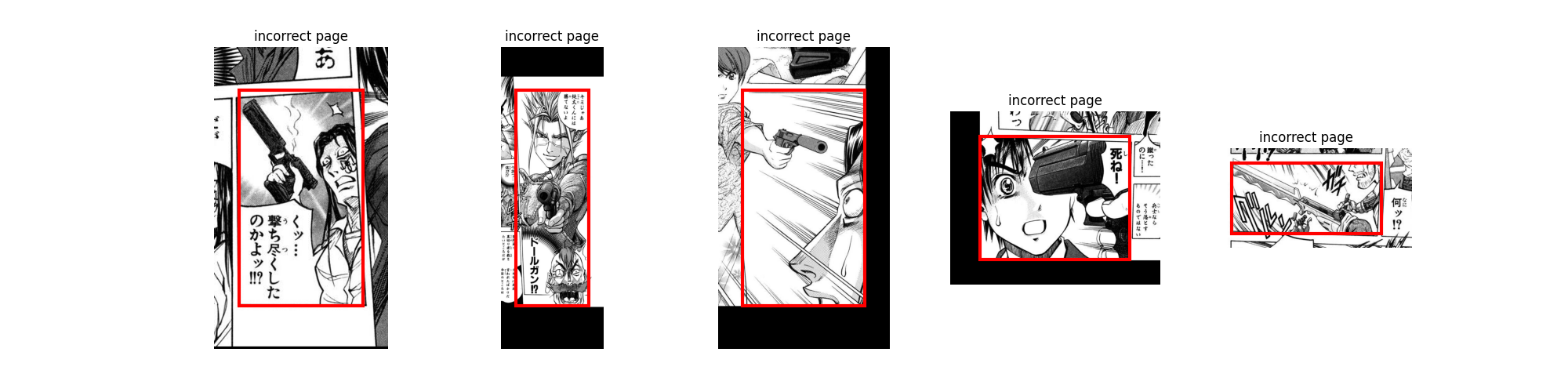}
    \caption{A man points a gun at his own head.}
    \label{fig:point-head}
  \end{subfigure}
  \caption{Visualizing the top 5 retrieved scenes for a set of sample queries, \copyright Deguchi Ryusei \cite{dollgun}}
  \label{fig:scene-desc-vis}
\end{figure*}





\clearpage
{\small
\bibliographystyle{ieee_fullname}
\bibliography{egbib}

\begin{thebibliography}{10}\itemsep=-1pt

\bibitem{multimedia_aizawa_2020_manga109}
Kiyoharu Aizawa, Azuma Fujimoto, Atsushi Otsubo, Toru Ogawa, Yusuke Matsui, Koki Tsubota, and Hikaru Ikuta.
\newblock Building a manga dataset ``manga109'' with annotations for multimedia applications.
\newblock {\em IEEE MultiMedia}, 27(2):8--18, 2020.

\bibitem{carion2020end}
Nicolas Carion, Francisco Massa, Gabriel Synnaeve, Nicolas Usunier, Alexander Kirillov, and Sergey Zagoruyko.
\newblock End-to-end object detection with transformers.
\newblock In {\em Computer Vision--ECCV 2020: 16th European Conference, Glasgow, UK, August 23--28, 2020, Proceedings, Part I 16}, pages 213--229. Springer, 2020.

\bibitem{chu2018text}
Wei-Ta Chu and Chih-Chi Yu.
\newblock Text detection in manga by deep region proposal, classification, and regression.
\newblock In {\em 2018 IEEE Visual Communications and Image Processing (VCIP)}, pages 1--4, 2018.

\bibitem{bert_japanese}
CL-Tohoku.
\newblock Bert pretrained on japanese texts.
\newblock \url{https://github.com/cl-tohoku/bert-japanese}, 2020.
\newblock Accessed on: June 4th, 2023.

\bibitem{dollgun}
Ryusho Deguchi.
\newblock {\em DollGun}, volume~2 of {\em DollGun}.
\newblock Akita Shoten, 2000.

\bibitem{devlin2019bert}
Jacob Devlin, Ming-Wei Chang, Kenton Lee, and Kristina Toutanova.
\newblock Bert: Pre-training of deep bidirectional transformers for language understanding, 2019.

\bibitem{dosovitskiy2020vit}
Alexey Dosovitskiy, Lucas Beyer, Alexander Kolesnikov, Dirk Weissenborn, Xiaohua Zhai, Thomas Unterthiner, Mostafa Dehghani, Matthias Minderer, Georg Heigold, Sylvain Gelly, Jakob Uszkoreit, and Neil Houlsby.
\newblock An image is worth 16x16 words: Transformers for image recognition at scale.
\newblock {\em CoRR}, abs/2010.11929, 2020.

\bibitem{huggingface2023}
Hugging Face.
\newblock Trainer api — transformers 4.12.0 documentation, 2023.

\bibitem{girdhar2023imagebind}
Rohit Girdhar, Alaaeldin El-Nouby, Zhuang Liu, Mannat Singh, Kalyan~Vasudev Alwala, Armand Joulin, and Ishan Misra.
\newblock Imagebind: One embedding space to bind them all, 2023.

\bibitem{he2015resnet}
Kaiming He, Xiangyu Zhang, Shaoqing Ren, and Jian Sun.
\newblock Deep residual learning for image recognition.
\newblock {\em CoRR}, abs/1512.03385, 2015.

\bibitem{jia2021scaling}
Chao Jia, Yinfei Yang, Ye Xia, Yi-Ting Chen, Zarana Parekh, Hieu Pham, Quoc~V. Le, Yunhsuan Sung, Zhen Li, and Tom Duerig.
\newblock Scaling up visual and vision-language representation learning with noisy text supervision, 2021.

\bibitem{johnson2019billion}
Jeff Johnson, Matthijs Douze, and Herv{\'e} J{\'e}gou.
\newblock Billion-scale similarity search with {GPUs}.
\newblock {\em IEEE Transactions on Big Data}, 7(3):535--547, 2019.

\bibitem{karpathy2015caption}
Andrej Karpathy and Li Fei-Fei.
\newblock Deep visual-semantic alignments for generating image descriptions, 2015.

\bibitem{manga-ocr}
W. Kha.
\newblock Manga ocr.
\newblock \url{https://github.com/kha-white/manga-ocr}, 2023.

\bibitem{khattab2022demonstrate}
Omar Khattab, Keshav Santhanam, Xiang~Lisa Li, David Hall, Percy Liang, Christopher Potts, and Matei Zaharia.
\newblock Demonstrate-search-predict: Composing retrieval and language models for knowledge-intensive {NLP}.
\newblock {\em arXiv preprint arXiv:2212.14024}, 2022.

\bibitem{kovanen-manga-order}
Samu Kovanen and Kiyoharu Aizawa.
\newblock A layered method for determining manga text bubble reading order.
\newblock In {\em 2015 IEEE International Conference on Image Processing (ICIP)}, pages 4283--4287, 2015.

\bibitem{li2022blip}
Junnan Li, Dongxu Li, Caiming Xiong, and Steven Hoi.
\newblock Blip: Bootstrapping language-image pre-training for unified vision-language understanding and generation, 2022.

\bibitem{mtap_matsui_2017}
Yusuke Matsui, Kota Ito, Yuji Aramaki, Azuma Fujimoto, Toru Ogawa, Toshihiko Yamasaki, and Kiyoharu Aizawa.
\newblock Sketch-based manga retrieval using manga109 dataset.
\newblock {\em Multimedia Tools and Applications}, 76(20):21811--21838, 2017.

\bibitem{NEURIPS2019_9015}
Adam Paszke, Sam Gross, Francisco Massa, Adam Lerer, James Bradbury, Gregory Chanan, Trevor Killeen, Zeming Lin, Natalia Gimelshein, Luca Antiga, Alban Desmaison, Andreas Kopf, Edward Yang, Zachary DeVito, Martin Raison, Alykhan Tejani, Sasank Chilamkurthy, Benoit Steiner, Lu Fang, Junjie Bai, and Soumith Chintala.
\newblock Pytorch: An imperative style, high-performance deep learning library.
\newblock In {\em Advances in Neural Information Processing Systems 32}, pages 8024--8035. Curran Associates, Inc., 2019.

\bibitem{radford2021clip}
Alec Radford, Jong~Wook Kim, Chris Hallacy, Aditya Ramesh, Gabriel Goh, Sandhini Agarwal, Girish Sastry, Amanda Askell, Pamela Mishkin, Jack Clark, Gretchen Krueger, and Ilya Sutskever.
\newblock Learning transferable visual models from natural language supervision, 2021.

\bibitem{reimers2019sentencebert}
Nils Reimers and Iryna Gurevych.
\newblock Sentence-bert: Sentence embeddings using siamese bert-networks, 2019.

\bibitem{reimer2020multilingual}
Nils Reimers and Iryna Gurevych.
\newblock Making monolingual sentence embeddings multilingual using knowledge distillation.
\newblock {\em CoRR}, abs/2004.09813, 2020.

\bibitem{ren2015faster}
Shaoqing Ren, Kaiming He, Ross Girshick, and Jian Sun.
\newblock Faster r-cnn: Towards real-time object detection with region proposal networks.
\newblock In {\em Proceedings of the 28th International Conference on Neural Information Processing Systems - Volume 1}, NIPS'15, page 91–99, Cambridge, MA, USA, 2015. MIT Press.

\bibitem{Rezatofighi_2018_CVPR}
Hamid Rezatofighi, Nathan Tsoi, JunYoung Gwak, Amir Sadeghian, Ian Reid, and Silvio Savarese.
\newblock Generalized intersection over union.
\newblock June 2019.

\bibitem{sanh2019distilbert}
Victor Sanh, Lysandre Debut, Julien Chaumond, and Thomas Wolf.
\newblock Distilbert, a distilled version of {BERT:} smaller, faster, cheaper and lighter.
\newblock {\em CoRR}, abs/1910.01108, 2019.

\bibitem{vit-clip}
SentenceTransformers.
\newblock \url{https://huggingface.co/sentence-transformers/clip-ViT-B-32}.
\newblock Accessed on: June 4th, 2023.

\bibitem{multilingual-clip}
SentenceTransformers.
\newblock \url{https://huggingface.co/sentence-transformers/clip-ViT-B-32-multilingual-v1}.
\newblock Accessed on: June 4th, 2023.

\bibitem{su2023pandagpt}
Yixuan Su, Tian Lan, Huayang Li, Jialu Xu, Yan Wang, and Deng Cai.
\newblock Pandagpt: One model to instruction-follow them all, 2023.

\bibitem{touvron2021deit}
Hugo Touvron, Matthieu Cord, Matthijs Douze, Francisco Massa, Alexandre Sablayrolles, and Hervé Jégou.
\newblock Training data-efficient image transformers \& distillation through attention, 2021.

\bibitem{vaswani2017transformer}
Ashish Vaswani, Noam Shazeer, Niki Parmar, Jakob Uszkoreit, Llion Jones, Aidan~N. Gomez, Lukasz Kaiser, and Illia Polosukhin.
\newblock Attention is all you need.
\newblock {\em CoRR}, abs/1706.03762, 2017.

\bibitem{voorhees1999proceedings}
Ellen~M. Voorhees.
\newblock Proceedings of the 8th text retrieval conference.
\newblock In {\em TREC-8 Question Answering Track Report}, pages 77--82, 1999.

\bibitem{zhu2023minigpt4}
Deyao Zhu, Jun Chen, Xiaoqian Shen, Xiang Li, and Mohamed Elhoseiny.
\newblock Minigpt-4: Enhancing vision-language understanding with advanced large language models, 2023.

\end{thebibliography}
}

\end{document}